\documentclass[sigconf,screen,nonacm]{acmart}
\newif\ifacmformat
\acmformattrue

\usepackage{mathtools}
\usepackage[nameinlink,capitalize,noabbrev]{cleveref}
\graphicspath{{figures/}}

\ifacmformat\theoremstyle{acmplain}\else\theoremstyle{plain}\fi
\newtheorem{theorem}{Theorem}[section]
\newtheorem{lemma}[theorem]{Lemma}
\newtheorem{corollary}[theorem]{Corollary}
\newtheorem{proposition}[theorem]{Proposition}

\ifacmformat\theoremstyle{acmdefinition}\else\theoremstyle{definition}\fi
\newtheorem{definition}[theorem]{Definition}

\crefname{target}{target statement}{target statements}
\Crefname{target}{Target Statement}{Target Statements}
\theoremstyle{remark}

\newcommand{\E}{\mathbb{E}}
\newcommand{\eps}{\varepsilon}
\newcommand{\Cfield}{\mathcal{C}_{\mathrm{field}}}

\DeclareMathOperator{\tr}{tr}

\newcommand{\Cstar}{\mathcal C_*}

\usepackage{orcidlink}
\newcommand{\PaperTitle}{Estimating a Transition Point from Gibbs and Ground States}
\newcommand{\PaperFirstAuthor}{Yuhao Cai}
\newcommand{\PaperFirstEmail}{SophiaCai1879@outlook.com}
\newcommand{\PaperFirstORCID}{0009-0004-4147-0997}
\newcommand{\PaperSecondAuthor}{Zhenzhang Li}
\newcommand{\PaperSecondEmail}{zhenzhangli@gpnu.edu.cn}
\newcommand{\PaperSecondORCID}{0000-0001-5504-7197}

\newcommand{\PaperInstitution}{Guangdong Polytechnic Normal University}
\newcommand{\PaperShenzhenInstitution}{Shenzhen Bay Laboratory}
\newcommand{\PaperXianInstitution}{Xi'an Jiaotong University}
\newcommand{\PaperXianLaboratoryName}{The Key Laboratory of Biomedical Information Engineering}
\newcommand{\PaperXianLaboratory}{\PaperXianLaboratoryName, Ministry of Education}
\newcommand{\PaperThirdAffiliation}{\PaperXianLaboratory, \mbox{\PaperXianInstitution}}
\newcommand{\PaperContactDetails}{%
  Corresponding author.\par
  \noindent\PaperFirstAuthor:
    \href{mailto:\PaperFirstEmail}{\texttt{\PaperFirstEmail}}.\par
  \noindent\PaperSecondAuthor:
    \href{mailto:\PaperSecondEmail}{\texttt{\PaperSecondEmail}}.}

\title{\PaperTitle}
\ifacmformat
  \author{\PaperFirstAuthor}
  \orcid{\PaperFirstORCID}
  \affiliation{%
    \institution{\PaperInstitution}
    \city{Guangzhou}
    \country{China}}
  \affiliation{%
    \institution{\PaperShenzhenInstitution}
    \city{Shenzhen}
    \country{China}}
  \email{\PaperFirstEmail}
  \author{\PaperSecondAuthor}
  \orcid{\PaperSecondORCID}
  \authornote{\PaperContactDetails}
  \affiliation{%
    \institution{\PaperInstitution}
    \city{Guangzhou}
    \country{China}}
  \affiliation{%
    \institution{\PaperThirdAffiliation}
    \city{Xi'an}
    \country{China}}
  \email{\PaperSecondEmail}
  \renewcommand{\shortauthors}{Cai and Li}

  \makeatletter
  \expandafter\def\csname typeset@author1\endcsname#1{%
    #1\nobreak\hspace{0.3em}\orcidlink{\PaperFirstORCID}}
  \expandafter\def\csname typeset@author2\endcsname#1{%
    #1\nobreak\hspace{0.3em}\orcidlink{\PaperSecondORCID}}
  \let\PaperNativeMakeAuthors\@mkauthors
  \renewcommand{\@mkauthors}{%
    \begingroup
      \def\city##1{\global\@ACM@citypresenttrue}%
      \def\country##1{\global\@ACM@countrypresenttrue}%
      \let\PaperNativeHref\href
      \renewcommand{\href}[2]{%
        \IfBeginWith{##1}{mailto:}{}{\PaperNativeHref{##1}{##2}}}%
      \PaperNativeMakeAuthors
    \endgroup}
  \makeatother
\else
  \author{%
    \begin{tabular}{@{}c@{\qquad}c@{}}
      \PaperFirstAuthor\textsuperscript{1,2} &
      \PaperSecondAuthor\textsuperscript{1,3,*}
    \end{tabular}\\[0.8em]
    \parbox{0.96\textwidth}{\centering\normalsize
      \textsuperscript{1}\PaperInstitution
      \\[0.5em]
      \textsuperscript{2}\PaperShenzhenInstitution
      \\[0.5em]
      \textsuperscript{3}\PaperXianLaboratoryName,\\
      Ministry of Education, \PaperXianInstitution}}
  \date{}
\fi

\setcopyright{none}
\acmDOI{}
\acmISBN{}
\keywords{quantum learning, Hamiltonian learning, phase transitions,
  sample complexity, finite-size effects}
\AtEndDocument{\nobalance}

\begin{document}
\begin{abstract}
  We study how the size of an observed many-body system affects the
number of state copies needed to estimate a zero-temperature
first-order transition point. The unknown Hamiltonian has bounded,
translation-invariant on-site and nearest-neighbour interactions on a
three-dimensional cubic lattice, with local dimension three. The field
coupling is known, and all models share a known boundary field that
penalizes one local level. The limiting order
parameter is uniformly separated on the two sides of the transition.
A query returns an independently prepared Gibbs or ground state at a
field and system size chosen adaptively by the learner.

For error $\eps$ and fixed confidence, we prove an exponential
separation between two size caps at the same temperature. There are
absolute constants $0<c_0<C$ and $c_1>0$ such that, at inverse
temperature $\beta=c_1/\eps$, restricting side lengths to
$L\le c_0/\eps$ requires $\exp(\Omega(1/\eps))$ copies in worst-case
expectation, even with collective measurements. Allowing
$L\le C/\eps$ permits $O(\log^2(1/\eps))$ copies and
$O(\eps^{-3})$ total spins, using local measurements and polynomial
classical postprocessing. The minimax expected total spin count is
$\Theta(\eps^{-3})$ uniformly for $\beta\ge c_1/\eps$.
At fixed positive temperature it is $\Theta_\beta(\eps^{-2})$,
attained from one-site and eight-site Gibbs states followed by
exponential classical computation. For ground-state copies it is
$\Theta(\eps^{-3})$.

The lower bound uses a three-state ferromagnet whose common boundary
suppresses the sector carrying information about the transition.
A spanning-forest argument and edge-boundary counting control its
Gibbs weight uniformly over fields and allowed sizes. A trace-distance
hybrid argument handles adaptive learners with quantum memory.
The upper bound combines finite-volume free-energy estimates,
concavity and local order-parameter measurements.

\end{abstract}
\maketitle
\section{Introduction}
\label{sec:introduction}
We study how large an observed many-body system must be to learn a
zero-temperature phase-transition point. The transition is defined as
the system size tends to infinity, whereas each observation comes from
a finite system. Boundary effects can shift its apparent
transition~\cite[full paper, Theorem~A, Eq.~(2.21a)]{BorgsKotecky1995},
and ground-state properties can change with
size~\cite[Main results~1--2]{BauschSize2018}. Repeated observations
improve knowledge of a fixed finite system. Increasing its size changes
the state being observed. We ask how these choices affect the sample
complexity of estimating the limiting transition.

\subsection{The learning problem}
The unknown interaction is bounded, translation invariant and
nearest neighbour on a three-dimensional cubic lattice with three
levels per site. A known field $\phi\in[0,1]$ couples linearly to a
known on-site observable. Its limiting expectation per site is the
zero-temperature order parameter. We assume a unique transition point
$\theta\in[1/4,3/4]$ and a uniform gap separating the order-parameter
values on its two sides. Every model uses the same known surface
field, which penalizes one local level at the boundary
(\cref{def:comparison-class}).

A query chooses a field and a side length $L$, and returns an
independently prepared equilibrium state of that cube: a Gibbs state
at known inverse temperature $\beta$, held fixed during the run, or
the normalized ground-space projector. The learner may choose queries
adaptively, retain quantum memory and measure received states jointly.
Its estimate must have error at most $\eps$ with the prescribed
probability, uniformly over the unknown interaction. We count copies,
maximum side length and total spins. State preparation is supplied by
the access model; classical processing is accounted for separately.

\subsection{Results and their significance}
Our main result compares two size caps at the same temperature.
For success probability $9/10$, let $\mathsf N_\beta^*(\eps;R)$ be the
smallest worst-case expected copy count among learners restricted to
$L\le R$. With $\kappa=1/16$, an absolute constant $C>\kappa$ gives
\begin{equation}\label{eq:intro-copy-separation}
 \begin{aligned}
 \mathsf N_{512/\eps}^*\!\left(\eps;
       \left\lfloor\frac\kappa\eps\right\rfloor\right)
     &\ge\frac45 e^{52/\eps},\\
 \mathsf N_{512/\eps}^*\!\left(\eps;
       \left\lceil\frac C\eps\right\rceil\right)
     &=O\!\left(\log^2\frac1\eps\right),
 \qquad 0<\eps\le1/16.
 \end{aligned}
\end{equation}
\Cref{eq:minimax-copies,cor:finite-size-separation} give the formal
definition and guarantee. The lower bound allows collective
measurements and almost-sure stopping. The upper learner uses local
measurements and polynomial classical processing in the accuracy input
length.

Both caps grow as $\eps^{-1}$. Increasing the permitted side length
by a constant factor therefore separates exponential from
polylogarithmic copy complexity, with the same temperature, field
range, boundary and unknown model class. System size is a separate
statistical parameter even with full measurement access to each copy.
The larger-system learner uses $O(\eps^{-3})$ total spins, which is
optimal uniformly for $\beta\ge512/\eps$ (\cref{thm:cold-minimax}).

At fixed positive temperature, the optimal total spin count is
$\Theta_\beta(\eps^{-2})$: one-site and eight-site Gibbs states suffice,
followed by exponential classical computation. For ground-state copies
it is $\Theta(\eps^{-3})$, and a worst-case learner must request side
length of order $\eps^{-1}$
(\cref{thm:temperature-separation,cor:temperature-cost}). The upper
bounds apply to noncommuting interactions. Diagonal interactions
establish the lower bounds against general quantum measurements.

\subsection{Technical overview}
\paragraph{Boundary effects.}
Our ferromagnet has local states $A,B,C$. Giving $C$ a bulk energy
advantage moves the thermodynamic transition, while the common
boundary penalizes $C$. The surface--volume competition is studied by
Borgs and Koteck\'y~\cite[full paper, Theorem~A]{BorgsKotecky1995}.
In our construction it leaves exactly the same ground states in small cubes,
at every field. \Cref{lem:whole-state-identity} gives these complete
density matrices; \cref{thm:ground-whole-state} gives the size lower bound.

\paragraph{Thermal information.}
At positive temperature the instances differ through configurations
containing $C$. Rare excited levels already give Hamiltonian-learning
lower bounds~\cite[Supplementary Information, Section~8]{HaahKothariTang2024};
low-temperature stability of a size-driven construction is analyzed
in~\cite[full paper, Eqs.~(7)--(10)]{BauschSize2018}.
The cited thermal lower bounds quantify information in a specified
Gibbs input. Here the learner controls both the field and the size.
The same pair of instances must remain close at every field and every
size up to $R=\Theta(\eps^{-1})$, at inverse temperature
$\Theta(\eps^{-1})$.

We retain the disagreement energy when summing configurations.
For a fixed nonempty $C$ set, root labels and agreement bits on a
spanning forest determine the $A/B$ labels of its complement.
The number of components is at most the edge boundary of the $C$ set,
which also determines the set itself. Summing boundaries gives an
excited-sector weight of order $R^{18}e^{-2\kappa\beta}$ when
$\kappa\beta\ge48\log(12(R+1))$ (\cref{lem:sector-probability}).
The polynomial size factor permits the copy lower bound even as the
allowed cubes grow with accuracy.

Conditioning on the absence of $C$ leaves the same state in both
instances. A trace-distance hybrid argument, following Bennett,
Bernstein, Brassard and Vazirani~\cite[full paper,
Theorem~3.3]{BennettHybrid1997}, sums the fresh-state differences
using actual query probabilities under either instance. Passing to
an almost-sure stopping time gives the spin and copy bounds in
\cref{lem:approximate-cost,cor:cold-copies}.

\paragraph{Free-energy comparisons.}
The upper bound uses the known field coupling. Free energy approximates
the limiting ground-energy density with errors of order $L^{-1}$ and
$\beta^{-1}$. The surface estimate follows by block tiling, as in
Aharonov and Irani~\cite[full paper,
Lemma~3.2]{AharonovIraniThermodynamic2025}. Concavity converts this
energy estimate into an order-parameter comparison away from the
transition. Local measurements and interval search implement the
learner. Assigning smaller failure probabilities to earlier, smaller
cubes gives $O(\log^2(1/\eps))$ copies and $O(\eps^{-3})$ total spins
(\cref{thm:cold-upper}).

\subsection{Related work}
Hamiltonian learning recovers the interaction of a finite system.
Bresler~\cite[full paper, Theorem~4.1]{BreslerIsing2015} gives efficient
structure learning for bounded-degree classical Ising models with fixed
interaction and field bounds. Anshu, Arunachalam, Kuwahara and
Soleimanifar~\cite[full paper, Theorem~2]{AnshuManyBody2021} establish
polynomial sample bounds for quantum Hamiltonian learning. Bakshi, Liu,
Moitra and Tang~\cite[full paper, Theorem~1.1 and
Remark~1.2]{BakshiHamiltonian2026} obtain polynomial running time at
every fixed positive temperature; Narayanan~\cite[Theorem~6]{Narayanan2025}
improves the temperature dependence. Our target is a transition in
the thermodynamic limit, with system size included in learning complexity.

Chen, Anshu and Nguyen~\cite[full paper, Theorem~I.2]{ChenAnshuNguyen2025}
learn from local measurements within a larger Gibbs state; our queries
prepare finite cubes independently. Gaitonde, Moitra and
Mossel~\cite[full paper, Theorem~1.1]{GaitondeDynamics2025} study the
role of dynamical observations in Markov-random-field learning.
Huang, Kueng, Torlai, Albert and Preskill~\cite[Theorems~1--2]{HuangManyBody2022}
predict properties in smooth, gapped families and classify phases with
few-body classifiers. Rouz\'e, Stilck Fran\c{c}a, Onorati and
Watson~\cite[Theorem~2]{RouzePhases2024} learn local observables in
thermal phases with exponential correlation decay. Classical
shadows~\cite[Theorem~1]{HuangShadows2020} estimate many observables
with guarantees depending on their shadow norms.

Classical interaction descriptions lead to a further computational
question. Cubitt, Perez-Garcia and Wolf~\cite[p.~207]{CubittSpectralGap2015}
prove spectral-gap undecidability. Watson and
Bausch~\cite[Definition~10 and Theorem~19]{WatsonBauschCritical2021}
study critical-parameter approximation under a local--global phase
promise. Watson and Cubitt~\cite[Definitions~2.2--2.3]{WatsonCubittGSED2022}
and Aharonov and Irani~\cite[full paper, Definition~1.2 and
Theorem~1.3]{AharonovIraniThermodynamic2025} fix the interaction and
study energy-density complexity with precision specified by the input.
\Cref{sec:computation} gives an approximation algorithm from rational
local matrices under our conjugate-field phase promise.

\section{The Estimation Problem}
\label{sec:preliminaries}
We define the unknown interaction, its limiting transition point, and
the finite-system observations available to the learner.

\subsection{Local interactions and the transition}
Let $\Lambda_L=\{1,\ldots,L\}^3$ and let the local Hilbert space be
$\mathbb C^q$. Fix a public algebraic operator $0\le P\le I$ and put
$M_L=\sum_{x\in\Lambda_L}P_x$. An instance $u$ specifies an on-site term $h_0$ on $\mathbb C^q$ and
three nearest-neighbour terms $h_i$ on $\mathbb C^q\otimes\mathbb C^q$.
The entries are algebraic, and their descriptions are unknown to the
learner. Write $e_i$ for the three coordinate unit vectors. Translation
invariance means that the same terms are used at every site and edge:
\begin{equation}\label{eq:local-interaction}
 K_{u,L}=\sum_{x\in\Lambda_L}(h_0)_x
       +\sum_{i=1}^3\sum_{x,x+e_i\in\Lambda_L}(h_i)_{x,x+e_i}.
\end{equation}
Every local term has operator norm at most a public rational $J$.
A public boundary prescription satisfies
$\|B_L\|\le J_bL^2$, where $J_b$ is rational, and has a uniform effective
algebraic encoding. The Hamiltonian is
\begin{equation}\label{eq:field-model}
 H_{u,L}(\phi)=K_{u,L}-\phi M_L+B_L,
 \qquad \phi\in[0,1].
\end{equation}
The limiting ground energy per site is denoted by $e_u(\phi)$.
\Cref{app:proofs} proves its existence and an explicit finite-size bound.

At zero temperature, we use the normalized projector onto the ground space:
\begin{equation}\label{eq:canonical-state}
 \rho^\infty_{u,L}(\phi)
 =\frac{\Pi_{\min}(H_{u,L}(\phi))}
        {\operatorname{tr}\Pi_{\min}(H_{u,L}(\phi))}.
\end{equation}
Write $\rho_{u,L}=\rho^\infty_{u,L}$ and
$m_{u,L}=L^{-3}\operatorname{tr}(\rho_{u,L}M_L)$.
An instance belongs to $\Cfield$ if there is a transition point
$\theta_u\in[1/4,3/4]$ and a limit $m_u(\phi)$ for every
$\phi\ne\theta_u$. Known rational numbers
$0\le a<b\le1$ give the promise
\begin{equation}\label{eq:phase-promise}
 m_u(\phi)\in
 \begin{cases}
 [0,a],&\phi<\theta_u,\\
 [b,1],&\phi>\theta_u.
 \end{cases}
\end{equation}
The gap $d=b-a$ is a lower bound on the order-parameter jump.
The promise holds throughout both field intervals. This uniform
separation determines the transition point and gives the comparisons
used by the upper learners.

\begin{definition}[The model class $\Cstar$]
\label{def:comparison-class}
Fix $q=3$, $P=\operatorname{diag}(0,1,1)$, $a=1/4$, $b=3/4$, $J=1$ and
$\kappa=1/16$. The class $\Cstar$ consists of the instances in
$\Cfield$ with these parameters and common boundary
\begin{equation}\label{eq:common-boundary}
 B_L=\kappa\sum_{x\in\Lambda_L}n_{\rm ext}(x)
                              |C\rangle\langle C|_x.
\end{equation}
Here $A,B,C$ label the public local basis, and $n_{\rm ext}(x)$ counts
edges from $x$ to the exterior. Thus $\|B_L\|\le6\kappa L^2$.
\end{definition}

\subsection{Access to Gibbs and ground states}
For a known rational $0<\beta<\infty$, held fixed throughout a run,
a Gibbs-state query returns
\begin{equation}\label{eq:thermal-state}
 \rho^\beta_{u,L}(\phi)
 =\frac{\exp[-\beta H_{u,L}(\phi)]}
        {\operatorname{tr}\exp[-\beta H_{u,L}(\phi)]}.
\end{equation}
This is the equilibrium state of the chosen finite Hamiltonian, with
the prescribed boundary. Local measurements within a larger Gibbs
state instead give its reduced states, as in Chen, Anshu and
Nguyen~\cite[full paper, Theorem~I.2]{ChenAnshuNguyen2025}.
For $\beta=\infty$, the query returns \cref{eq:canonical-state}.
In each case the query is a dyadic $\phi\in[0,1]$ and an integer
$L\ge1$, and returns a fresh, independent copy. The learner may retain
quantum memory and apply instance-independent quantum instruments,
including joint measurements of all received copies. The algorithm chooses each field value and system size classically,
using its preceding measurement outcomes.

Accuracy and confidence inputs are positive rational numbers.
An algorithm terminates almost surely on every instance and returns a
rational $\widehat\theta\in[1/4,3/4]$. The transcript includes all
classical queries, instrument outcomes, the stopping decision and the
final output. Write $N$ for the number of copies and $L_{\max}$ for
their maximum side length, with $L_{\max}=0$ when $N=0$.
The total number of spins in the copies is
\begin{equation}\label{eq:resources}
 V=\sum_{t=1}^{N}L_t^3.
\end{equation}
Let $\mathfrak A^\beta_{\eps,\delta}(\mathcal C)$ contain the algorithms
satisfying
\begin{equation}\label{eq:success}
 \sup_{u\in\mathcal C}
 \Pr_u^{\mathsf A}\{|\widehat\theta-\theta_u|>\eps\}\le\delta.
\end{equation}
For fixed $\delta_*=1/10$, define
\begin{equation}\label{eq:minimax-cost}
 \mathsf V_\beta^*(\eps)
 =\inf_{\mathsf A\in\mathfrak A^\beta_{\eps,\delta_*}(\Cstar)}
       \sup_{u\in\Cstar}\mathbb E_u^{\mathsf A}[V].
\end{equation}
For an integer side-length cap $R\ge1$, let
$\mathfrak A^\beta_{\eps,\delta_*}(\Cstar;R)$ consist of the algorithms
in $\mathfrak A^\beta_{\eps,\delta_*}(\Cstar)$ satisfying
$L_{\max}\le R$ almost surely under every $u\in\Cstar$.
The corresponding copy complexity is
\begin{equation}\label{eq:minimax-copies}
 \mathsf N_\beta^*(\eps;R)
 =\inf_{\mathsf A\in\mathfrak A^\beta_{\eps,\delta_*}(\Cstar;R)}
       \sup_{u\in\Cstar}\mathbb E_u^{\mathsf A}[N].
\end{equation}
Preparation of the requested copies is supplied by the access model,
as in Gibbs-copy Hamiltonian
learning~\cite[full paper, Problem~1]{AnshuManyBody2021}.
Classical running time measures processing after receiving these
inputs. We use $\inf\varnothing=\infty$ in the minimax definitions.

\subsection{A local measurement}
The ground-state algorithm measures $\{P,I-P\}$ at one uniformly
random site of each copy. The outcome is a bit $Y$ with distribution
\begin{equation}\label{eq:measurement-law}
 Y\sim\operatorname{Ber}(m_{u,L}(\phi)).
\end{equation}
Independent copies give independent outcomes. This measurement is
allowed by both access models; the lower bounds allow collective
measurements as well.

\section{System Size and Sample Complexity}
\label{sec:comparison}
The main theorem compares two size caps on the same class $\Cstar$,
at the same temperature. Correctness is uniform over the unknown
interaction; the boundary is the common prescription in
\cref{def:comparison-class}.

\begin{theorem}[System size and number of copies]
\label{cor:finite-size-separation}
Let $0<\eps\le1/16$, $\beta=512/\eps$ and $\delta=1/10$.
Every $(\eps,\delta)$ learner on $\Cstar$ restricted to
$L\le\lfloor\kappa/\eps\rfloor$ has worst-case expected copy
count at least $(4/5)e^{52/\eps}$.
There is an absolute constant $C>\kappa$ and an $(\eps,\delta)$ learner
with $L_{\max}\le\lceil C/\eps\rceil$, using
$O(\log^2(1/\eps))$ copies and $O(\eps^{-3})$ total spins.
It uses local measurements and expected classical processing time
polynomial in the accuracy input bit length.
\end{theorem}
\begin{proof}
The two-instance bound in \cref{cor:cold-copies} applies to every
learner correct on $\Cstar$. Substituting $\beta=512/\eps$ and
$\delta=1/10$ in \cref{eq:cold-copy-lower} gives the lower bound,
since $13\kappa\cdot512/8=52$.
For the upper bound, apply \cref{thm:cold-upper} with $d=1/2$.
Its size constant depends only on the fixed public parameters of
$\Cstar$, so an absolute $C$ gives the stated cap. The same theorem
gives the copy, spin and classical processing bounds.
\end{proof}

\begin{theorem}[Temperature regimes]\label{thm:temperature-separation}
Let $0<\eps\le1/16$, and require error probability at most $1/10$
uniformly over $\Cstar$.
\begin{enumerate}
\item For every fixed rational $\beta>0$, there is an algorithm using
$O_\beta(\eps^{-2})$ copies of the Gibbs states of $H_{u,1}(0)$ and
$H_{u,2}(0)$. If an algorithm uses only systems with $L\le2$, then its
worst-case expected number of copies is $\Omega_\beta(\eps^{-2})$,
even with collective measurements.
\item With ground-state copies, there are two instances such that every
correct algorithm satisfies, under each instance,
\begin{equation}\label{eq:large-state-probability}
 \Pr\!\left\{L_{\max}>\left\lfloor\frac{\kappa}{\eps}\right\rfloor\right\}
 \ge\frac45.
\end{equation}
There is an algorithm with $L_{\max}=O(\eps^{-1})$ and
$O(\log^2(1/\eps))$ copies.
\item There are constants $c,C>0$ independent of $\eps,\beta$ such
that, for every rational $\beta\ge512/\eps$,
\begin{equation}\label{eq:main-cold-rate}
 c\eps^{-3}\le\mathsf V_\beta^*(\eps)\le C\eps^{-3}.
\end{equation}
The upper learner uses $L_{\max}=O(\eps^{-1})$ and
$O(\log^2(1/\eps))$ copies.
\end{enumerate}
The lower bounds allow adaptive measurements and quantum memory.
\end{theorem}

At fixed positive temperature, copies of systems with at most eight
spins suffice. At zero temperature, the two hard instances have
identical states below the stated size threshold. The distinction
holds for the complete density matrices supplied by a query.

For the cost $V=\sum_tL_t^3$ defined in \cref{eq:resources}, the bounds
also give the following minimax statement.
\begin{corollary}[Minimax total spin count]\label{cor:temperature-cost}
On the same class $\Cstar$ and at the same confidence,
\begin{equation}\label{eq:temperature-rate}
 \mathsf V_\beta^*(\eps)=
 \begin{cases}
 \Theta_\beta(\eps^{-2}),&0<\beta<\infty\text{ fixed},\\
 \Theta(\eps^{-3}),&\beta=\infty.
 \end{cases}
\end{equation}
\end{corollary}

\begin{proof}[Proof of \cref{thm:temperature-separation,cor:temperature-cost}]
The Gibbs-state algorithm is given in \cref{thm:thermal-upper}.
\Cref{thm:thermal-lower} proves a lower bound on $\mathbb E[V]$.
When $L\le2$, $V\le8N$, so it also proves the stated sample lower bound.
\Cref{thm:ground-whole-state} proves the ground-state size bound and
$\mathbb E[V]=\Omega(\eps^{-3})$.
\Cref{thm:cold-minimax} proves the third part uniformly in the stated
finite-temperature regime.

For the ground-state upper bound, the surface-energy estimate is
$|e_{u,L}-e_u|\le C/L$. Concavity of the energy makes
$-m_{u,L}(\phi)$ a supergradient. At distance $\gamma$ from the
transition, the resulting error in the order-parameter bounds is at
most $4C/(L\gamma)$. Taking $L=\lceil16C/(d\gamma)\rceil$ makes a local
measurement distinguish the two sides with a constant gap. Interval
search with overlapping updates contracts the uncertainty interval by
$5/8$. A geometric choice of round failure probabilities gives
$O(\log^2(1/\eps))$ copies and total cost $O(\eps^{-3})$ at fixed
confidence. \Cref{sec:main-results,app:proofs} gives the full algorithm
and cost calculation.
\end{proof}

\subsection{Interval search}
\label{subsec:interval-search}
All three upper bounds reduce the estimate to comparisons away from
the transition. Within the comparison margin, either answer is allowed;
overlapping interval updates preserve the target in both cases.
The measurement and reconstruction arguments establish the comparison
condition separately. Start with
$I_0=[1/4,3/4]$. At round $k$, write $I_k=[\ell_k,r_k]$,
let $\phi_k$ be its midpoint and $w_k$ its width, and put
$\gamma_k=w_k/8$. A comparison returns a bit $Z_k$. Update by
\begin{equation}\label{eq:buffered-update}
 I_{k+1}=\begin{cases}
 [\ell_k,\phi_k+\gamma_k],&Z_k=1,\\
 [\phi_k-\gamma_k,r_k],&Z_k=0.
 \end{cases}
\end{equation}
The two intervals overlap on the points within $\gamma_k$ of $\phi_k$.
Stop after $K=\lceil\log(1/(4\eps))/\log(8/5)\rceil$ rounds and
output the midpoint of $I_K$.

\begin{lemma}[Interval invariant]\label{lem:interval-invariant}
Suppose each comparison gives $Z_k=0$ when
$\phi_k<\theta-\gamma_k$ and $Z_k=1$ when
$\phi_k>\theta+\gamma_k$. Then every $I_k$ contains $\theta$,
and the output has error at most $\eps$.
\end{lemma}
\begin{proof}
If $Z_k=1$, the comparison condition implies
$\theta\le\phi_k+\gamma_k$; if $Z_k=0$, it implies
$\theta\ge\phi_k-\gamma_k$. Thus either update preserves containment.
Its width is $5w_k/8$, so $w_K\le2\eps$.
\end{proof}

\section{A Ferromagnetic Construction}
\label{sec:lower-comparison}
We construct two different thermodynamic transition points with
identical ground-state answers below a size threshold. A boundary
penalty competes with a bulk energy advantage. Surface-induced
transition shifts~\cite[full paper, Theorem~A]{BorgsKotecky1995} and size-driven
ground-state changes~\cite[Main results~1--2]{BauschSize2018} are related
finite-size phenomena. We give the interaction and exact finite
ground spaces needed for the statistical lower bound.

\subsection{Identical small-volume ground states}
The construction uses the following cubic edge bound.
\begin{lemma}[Edge boundary of a subset of the cube]
\label{lem:grid-perimeter}
For every nonempty $S\subseteq\Lambda_L$, let $\partial S$ be the
nearest-neighbour edges from $S$ to its complement in the infinite
cubic lattice. Then
\begin{equation}\label{eq:cubic-perimeter}
 |\partial S|\ge\frac{6|S|}{L}.
\end{equation}
Equality holds precisely when $S=\Lambda_L$.
\end{lemma}
\begin{proof}
Let $n_i$ be the number of coordinate lines in direction $i$ that
meet $S$. Each contains at most $L$ points, so $n_iL\ge|S|$.
Each such line contributes at least two edges to $\partial S$.
Summing over the three directions proves the inequality.
At equality, every occupied coordinate line is full. Starting at
one point of $S$ and changing its coordinates successively reaches
every point of the cube, so $S=\Lambda_L$.
\end{proof}

Let $\omega_{s,L}=|s\rangle\langle s|^{\otimes L^3}$.
Consider a three-state ferromagnet with $P=\operatorname{diag}(0,1,1)$,
on-site term $h_0=\operatorname{diag}(0,1/2,1/2-r)$, and the same
interaction in each direction:
\begin{equation}\label{eq:ferromagnetic-edge}
 h_i=\kappa\left(I-\sum_{s\in\{A,B,C\}}|ss\rangle\langle ss|\right),
 \qquad i=1,2,3.
\end{equation}
Thus disagreements cost $\kappa$. Use the boundary term in
\cref{eq:common-boundary}.
For a basis configuration $s\in\{A,B,C\}^{\Lambda_L}$, write
$S_C=\{x:s_x=C\}$, and let $e_{AB}$ count internal edges whose endpoints
are $A$ and $B$. The boundary $\partial S_C$ includes exterior edges.

\begin{lemma}[Ground states at small system sizes]
\label{lem:whole-state-identity}
For $0\le r\le1/4$ and $rL<6\kappa$,
\begin{equation}\label{eq:whole-state-identity}
 \rho^\infty_{r,L}(\phi)=
 \begin{cases}
 \omega_{A,L},&\phi<1/2,\\
 (\omega_{A,L}+\omega_{B,L})/2,&\phi=1/2,\\
 \omega_{B,L},&\phi>1/2.
 \end{cases}
\end{equation}
This model belongs to $\Cstar$, with transition point
$\theta_r=1/2-r$.
\end{lemma}
\begin{proof}
Write $h=1/2-\phi$.
If $h>0$, replace every $B$ by $A$. The remaining $A/C$ configuration
has energy $(h-r)|S_C|+\kappa|\partial S_C|$, bounded below by
$(h-r+6\kappa/L)|S_C|$. The coefficient is positive, so all $A$ is
unique. If $h<0$, replacing $A$ by $B$ gives
$hL^3-r|S_C|+\kappa|\partial S_C|$; again $6\kappa/L-r>0$, and all
$B$ is unique. These uses of the cubic edge bound are justified in
\cref{lem:grid-perimeter}.

At $h=0$, let $e_{AB}$ count internal $A/B$ disagreement edges.
The exact configuration energy is
\begin{equation}\label{eq:zero-field-potts-energy}
 E=\kappa|\partial S_C|-r|S_C|+\kappa e_{AB}.
\end{equation}
It is at least $(6\kappa/L-r)|S_C|+\kappa e_{AB}$. Thus a ground
configuration has no $C$ sites and no $A/B$ edge. The cube graph is
connected, so it is all $A$ or all $B$. The normalized projector equally
weights these two configurations.

All edge and boundary terms are nonnegative. Therefore the energy per
site is at least the least on-site energy, $\min\{0,h-r\}$.
The uniform $A$ and $C$ configurations attain this bound to within
$6\kappa/L$.
Consequently
\begin{equation}\label{eq:hard-family-energy}
 e_r(\phi)=\min\{0,1/2-r-\phi\}.
\end{equation}
At every $\phi\ne1/2-r$, convergence of the energies and the
supergradient inequalities force $m_{r,L}(\phi)\to-e_r'(\phi)$.
The limiting order parameter is zero on the left and one on the
right, satisfying \cref{eq:phase-promise}. The local norms obey the
bounds of $\Cstar$ for $0\le r\le1/4$.
\end{proof}

\begin{lemma}[Indistinguishability under adaptive measurements]
\label{lem:stopped-state-equality}
Suppose two instances return identical states for every field and every
$L\le R$. For any algorithm, the output subprobability
measure on termination before requesting $L>R$ is identical under the
two instances. Their probabilities of issuing a first such query are
also identical.
\end{lemma}
\begin{proof}
Before the first larger query, the joint state of the classical history
and quantum memory is identical under the two instances. A query adds
the same state conditional on the history, and the same quantum
instrument preserves equality. This induction also includes termination,
the reported estimate, and the decision to request a larger system.
Taking the increasing union over finite histories proves the claim.
\end{proof}

Equality in \cref{eq:whole-state-identity} concerns the full density
matrices supplied by a query. The next theorem therefore applies to
every measurement of those matrices, including joint measurements of
states obtained at different fields and sizes.

\begin{theorem}[Ground-state size lower bound]
\label{thm:ground-whole-state}
For $0<\eps\le1/16$ there are two instances of $\Cstar$ such that every
algorithm with error at most $\delta<1/2$ on both obeys
the following bounds. Put $R=\lfloor\kappa/\eps\rfloor$. Then
\begin{equation}\label{eq:ground-size-lower}
 \begin{aligned}
 \Pr\{L_{\max}>R\}&\ge1-2\delta,\\
 \mathbb E[V]&\ge(1-2\delta)(R+1)^3,
 \end{aligned}
\end{equation}
The inequalities hold under each instance.
\end{theorem}
\begin{proof}
Choose rational $\eps'\in(\eps/2,\eps]$ and take $r_0=0$,
$r_1=4\eps'$. The targets are more than $2\eps$ apart, and
$r_1L\le4\kappa<6\kappa$ for $L\le R$.
\Cref{lem:whole-state-identity} gives equality of the complete states
at every such query. Let $p$ be the common probability of requesting
$L>R$. By almost-sure termination and
\cref{lem:stopped-state-equality}, the common stopping output measure
before that query has mass $1-p$. Its contributions to the two disjoint
success intervals sum to at most $1-p$. Success after a larger query
contributes at most $p$ under each instance. Hence the sum of the two
success probabilities is at most $1+p$, and correctness makes it at
least $2(1-\delta)$. This proves $p\ge1-2\delta$. Every run in that
event uses a copy of volume at least $(R+1)^3$.
\end{proof}

\section{Temperature and System Size}
\label{sec:finite-temperature}
We prove the upper bound on all of $\Cstar$, then analyze the
ferromagnetic pair from \cref{sec:lower-comparison}. A uniform bound
on its excited-sector probability gives the lower bound for adaptive
learners with quantum memory.

\subsection{Local measurements at low temperature}
Log-partition convexity is used in Hamiltonian identification by Anshu,
Arunachalam, Kuwahara and Soleimanifar~\cite[full paper,
Theorem~6]{AnshuManyBody2021}. Here the varying parameter is the known
conjugate field. We derive the one-parameter free-energy identity and
the approximation error used in the comparison algorithm.
Let $C=\max\{1,3J+J_b\}$, and define the finite-volume free energy
per site by
\begin{equation}\label{eq:finite-free-energy}
 f_{\beta,L}(\phi)=-\frac{1}{\beta L^3}
             \log\operatorname{tr}e^{-\beta H_{u,L}(\phi)}.
\end{equation}
Tile a larger cube by $L$-cubes and remove edges between blocks.
Their energy contribution per site is at most $3J/L$; the boundary
adds at most $J_b/L$. Thus the finite ground-energy density differs
from $e_u$ by at most $C/L$ (\cref{eq:surface-bound}). The partition
function lies between its largest summand and $3^{L^3}$ times that
summand, giving
\begin{equation}\label{eq:free-energy-error}
 |f_{\beta,L}(\phi)-e_u(\phi)|
 \le\frac{C}{L}+\frac{\log3}{\beta}.
\end{equation}
For block tiling, see~\cite[full paper,
Lemma~3.2]{AharonovIraniThermodynamic2025};
\cref{app:proofs} gives the full three-dimensional count.
Termwise differentiation of the exponential series and cyclicity of
trace give $f'_{\beta,L}=-m^\beta_{u,L}$, where
$m^\beta_{u,L}=L^{-3}\operatorname{tr}(\rho^\beta_{u,L}M_L)$.
This function is concave. Indeed, put
$\Delta M=M_L-\operatorname{tr}(\rho^\beta_{u,L}M_L)I$ and abbreviate
$\rho=\rho^\beta_{u,L}$. Differentiation gives
\begin{equation}\label{eq:free-energy-concavity}
 f''_{\beta,L}(\phi)
 =-\frac{\beta}{L^3}\int_0^1
       \operatorname{tr}(\rho^s\Delta M\rho^{1-s}\Delta M)\,ds
 \le0.
\end{equation}
Put $A_s=\rho^{(1-s)/2}\Delta M\rho^{s/2}$.
The integrand is $\operatorname{tr}(A_s^*A_s)\ge0$.

For finite ground states, the variational principle makes
$-m_{u,L}(\phi)$ a supergradient of $e_{u,L}$.
Passing to the limit and integrating the phase promise bounds the
secant slopes of $e_u$ by $[-a,0]$ to the left of $\theta_u$ and
by $[-1,-b]$ to its right.
Let $\eta=C/L+\log3/\beta$ and $h=\gamma/2$.
If $\phi\le\theta_u-\gamma$, concavity and
\cref{eq:free-energy-error} give
\begin{equation}\label{eq:thermal-secant-comparison}
 \begin{aligned}
 -m^\beta_{u,L}(\phi)
 &=f'_{\beta,L}(\phi)\\
 &\ge\frac{f_{\beta,L}(\phi+h)-f_{\beta,L}(\phi)}{h}\\
 &\ge\frac{e_u(\phi+h)-e_u(\phi)}{h}-\frac{2\eta}{h}\\
 &\ge-a-\frac{4\eta}{\gamma}.
 \end{aligned}
\end{equation}
For $\phi\ge\theta_u+\gamma$, the backward secant gives
$m^\beta_{u,L}(\phi)\ge b-4\eta/\gamma$.
These comparisons hold throughout the control interval.

\begin{theorem}[Learning from larger Gibbs states]
\label{thm:cold-upper}
Let $0<\eps\le1/16$, $0<\delta\le1/4$ and
$\beta\ge256/(d\eps)$. There is an $(\eps,\delta)$ learner on
$\Cstar$ with $L_{\max}=O(\eps^{-1})$ and
\begin{equation}\label{eq:cold-upper}
 V=O\!\left(\eps^{-3}\log\frac1\delta\right).
\end{equation}
It uses $O(K\log(1/\delta)+K^2)$ copies, where
$K=O(\log(1/\eps))$, and local measurements only.
\end{theorem}
\begin{proof}
At round $k$ of \cref{subsec:interval-search}, choose
$L_k=\lceil32C/(d\gamma_k)\rceil$.
Then $C/L_k\le d\gamma_k/32$.
Since $\log3<2$ and every executed $\gamma_k>\eps/4$,
$\log3/\beta\le d\eps/128<d\gamma_k/32$.
The secant comparisons separate the mean from $(a+b)/2$ by at
least $g=d/4$. Let $K$ be the number of rounds in
\cref{subsec:interval-search}, and allocate
$\delta_k=\delta2^{k-K}$ to round $k$. Take an integer number $n_k$
of copies with
\begin{equation}\label{eq:cold-round-samples}
 2g^{-2}\log(2/\delta_k)\le n_k
 <2g^{-2}\log(2/\delta_k)+2.
\end{equation}
A rational upper approximation to the logarithm selects this integer.
On each copy, measure $P$ at a uniformly random site. Set $Z_k=1$
if the sample mean is at least $(a+b)/2$, and set $Z_k=0$ otherwise.
Conditional on each history, Hoeffding's
inequality~\cite[Theorem~2, Eq.~(2.6)]{Hoeffding1963} bounds a
deviation of $g/2$ by $\delta_k$. These probabilities sum to less
than $\delta$, so \cref{lem:interval-invariant} proves correctness.
The sample counts have the form
$O(d^{-2}[\log(1/\delta)+K-k])$; summing them gives the copy bound.
The volumes grow by a geometric factor, so summing $n_kL_k^3$ gives
\cref{eq:cold-upper}; \cref{app:spin-count} supplies the geometric
calculation with the ground-state side-length constant.
The classical processing uses random-site
generation, count updates and rational interval arithmetic, and is
polynomial in expectation in the accuracy and confidence input bit lengths.
\end{proof}

\subsection{Information in small Gibbs states}
The ground-state identity becomes an approximate identity at positive
temperature. The information loss from rare excited levels underlies
the coefficient-learning bound of Haah, Kothari and
Tang~\cite[Supplementary Information, Section~8]{HaahKothariTang2024}.
Our estimate must hold simultaneously for every side length allowed by
the cap and every field the learner can select.
Put $R=\lfloor\kappa/\eps\rfloor$ and use the ferromagnetic pair
$r_0=0$, $r_1=4\eps'$, where rational
$\eps'\in(\eps/2,\eps]$. For $L\le R$ and $r\le4\eps$,
$rL\le4\kappa$. The cubic edge bound gives
$r|S_C|\le(2\kappa/3)|\partial S_C|$.
Write $h=1/2-\phi$, let $n_+(s)$ count the $B$ and $C$ sites, and
put $g=\min\{0,hL^3\}$. Since $hn_+(s)\ge g$, the energy obeys
\begin{equation}\label{eq:sector-weight-energy}
 \begin{aligned}
 E_r(s)-g
 &=hn_+(s)-g-r|S_C|+\kappa|\partial S_C|+\kappa e_{AB}\\
 &\ge\frac\kappa3|\partial S_C|+\kappa e_{AB}.
 \end{aligned}
\end{equation}
We retain the disagreement energy in this estimate to count the
excited configurations.

\begin{lemma}[Probability of the excited sector]
\label{lem:sector-probability}
Under either instance, at every field and every $L\le R$, the
probability of a configuration containing $C$ is at most
\begin{equation}\label{eq:small-gibbs-error}
 \begin{aligned}
 \xi_{\beta,R}
 &=\frac{(12R^3)^6}{6!}e^{-2\kappa\beta}\\
 &\quad\times\exp\!\left(R^3e^{-\kappa\beta}
                       +12R^3e^{-\kappa\beta/3}\right).
 \end{aligned}
\end{equation}
If $\kappa\beta\ge48\log(12(R+1))$, then
$\xi_{\beta,R}\le(R+1)^{-3}$.
\end{lemma}
\begin{proof}
Fix a nonempty $C$ set $S$. On the induced graph of its complement,
choose a spanning forest with $c$ components. Root labels and the
agreement bits along its edges determine all $A/B$ labels. Hence
\begin{equation}\label{eq:forest-partition}
 \sum_{A/B\text{ labels}}e^{-\beta\kappa e_{AB}}
 \le2^c(1+e^{-\beta\kappa})^{L^3-|S|-c}
 \le2^{|\partial S|}e^{L^3e^{-\beta\kappa}}.
\end{equation}
For the last inequality, every complement component touches $S$
because the cube is connected; thus $c\le|\partial S|$.
The empty complement has $c=0$ and satisfies the same bound.

The boundary map $S\mapsto\partial S$ is injective: equality of the
boundaries makes the two membership indicators have the same changes
on every edge, and their common exterior value is zero.
There are $M=3L^3+3L^2\le6L^3$ possible boundary edges, and every
nonempty set has at least six boundary edges. With
$a=2e^{-\beta\kappa/3}$, the total excited Boltzmann weight relative
to the reference is at most
\begin{equation}\label{eq:boundary-count-tail}
 e^{L^3e^{-\beta\kappa}}
 \sum_{b=6}^{M}\binom Mb a^b
 \le e^{L^3e^{-\beta\kappa}}\frac{(Ma)^6e^{Ma}}{6!}.
\end{equation}
Here $\binom Mb\le M^b/b!$ and
$b!\ge6!(b-6)!$ bound the tail. The partition function is at least
the reference weight, so substituting $M\le6L^3$ and $L\le R$
gives \cref{eq:small-gibbs-error}.

For the final assertion put $A=12(R+1)$. The assumed temperature gives
$e^{-\beta\kappa/3}\le A^{-16}$.
Both $R^3e^{-\beta\kappa}$ and
$12R^3e^{-\beta\kappa/3}$ are at most one.
Also $(12R^3)^6\le A^{18}$ and
$e^{-2\beta\kappa}\le A^{-96}$. Since $e^2/6!<1$,
$\xi_{\beta,R}\le A^{-78}\le(R+1)^{-3}$.
\end{proof}

Low-temperature stability of a hidden sector is analyzed by Bausch,
Cubitt, Lucia, Perez-Garcia and
Wolf~\cite[full paper, Eqs.~(7)--(10)]{BauschSize2018}.
The count above retains the $A/B$ disagreement energy and will enter
an adaptive estimation bound.

Conditioning on the absence of $C$ gives the same density matrix
$\sigma_{AB}$ under both instances. Put $p=\min\{1,\xi_{\beta,R}\}$.
Each Gibbs state contains the positive operator $(1-p)\sigma_{AB}$.
The remaining positive operators each have trace $p$, so the two
states have trace distance at most $p\le\xi_{\beta,R}$.
We use the hybrid argument to accumulate these distances along an
adaptive computation; the unitary-oracle form appears in Bennett,
Bernstein, Brassard and Vazirani~\cite[full paper, Theorem~3.3]{BennettHybrid1997}. The following proof gives the
fresh-state version, including the random stopping time.

\begin{samepage}
\begin{lemma}[Cost bound from approximate equality]
\label{lem:approximate-cost}
For the two instances just defined, every algorithm with error at most
$\delta<1/2$ on both satisfies, under either one,
\begin{equation}\label{eq:approximate-cost}
 \mathbb E[V]\ge
 \frac{1-2\delta}{\xi_{\beta,R}+(R+1)^{-3}}.
\end{equation}
The bound allows quantum memory and collective measurements.
Measurements and stopping decisions may be adaptive.
\end{lemma}
\end{samepage}
\begin{proof}
Write $D(\sigma,\tau)=\|\sigma-\tau\|_1/2$ for trace distance,
also using $D$ for total variation of classical laws. A query
$q=(\phi,L)$ appends the fresh state $\rho_i(q)$ under instance $i$.
Set $\eta(q)=\xi_{\beta,R}$ for $L\le R$ and $\eta(q)=1$ otherwise.
Then $D(\rho_0(q),\rho_1(q))\le\eta(q)$ at every query.

Pad a terminated run with an absorbing output register and dummy
queries whose two answers agree and whose $\eta$ is zero.
Let $\Sigma_i^t$ be the joint state of all classical registers and
quantum memory after round $t$. The initial state is common.
Just before the answer in round $t$, the classical query register makes the
state block diagonal. Under reference instance $i$, write its
unnormalized memory block at active query $q$ as $W_i^t(q)$.
Changing only that answer produces distance
\begin{equation}\label{eq:hybrid-query-increment}
 \begin{aligned}
 &\frac12\left\|
   \bigoplus_q W_i^t(q)\otimes
        (\rho_0(q)-\rho_1(q))\right\|_1\\
 &\qquad=\sum_q\tr W_i^t(q)D(\rho_0(q),\rho_1(q))\\
 &\qquad\le\mathbb E_i[\mathbf1_{\{N\ge t\}}\eta(Q_t)].
 \end{aligned}
\end{equation}
Here $Q_t$ is the $t$th active query. Positivity of each memory block
gives the tensor-norm identity; its trace is the actual query
probability under $i$.
The countable block sums converge absolutely, since the active block
traces sum to at most one.

Common quantum operations contract the trace norm of Hermitian
differences: write $X=X_+-X_-$, use positivity and trace preservation,
and apply the triangle inequality. Apply this contraction to the
common operations before and after the answer, then add and subtract
the state obtained by changing the oracle on the reference input.
For either $i$, \cref{eq:hybrid-query-increment} gives
\begin{equation}\label{eq:hybrid-round-recursion}
 D(\Sigma_0^t,\Sigma_1^t)
 \le D(\Sigma_0^{t-1},\Sigma_1^{t-1})
       +\mathbb E_i[\mathbf1_{\{N\ge t\}}\eta(Q_t)].
\end{equation}
This includes every adaptive measurement through the common
operations and their retained classical outcomes.

At horizon $T$, read the stored output or a timeout symbol.
Its distance is at most the sum of the increments through $T$.
Under each instance this output law converges in total variation to
the final law $P_i$, since $\Pr_i(N>T)\to0$.
Monotone convergence therefore gives
\begin{equation}\label{eq:hybrid-stopped-output}
 D(P_0,P_1)
 \le\mathbb E_i\!\left[\sum_{t=1}^N\eta(Q_t)\right]
 \le\xi_{\beta,R}\mathbb E_i[N]
       +\frac{\mathbb E_i[V]}{(R+1)^3}.
\end{equation}
The last inequality charges each query with $L>R$ at least
$(R+1)^3$ spins. The two disjoint success intervals imply
$D(P_0,P_1)\ge1-2\delta$. Finally $N\le V$ proves
\cref{eq:approximate-cost} under each instance.
\end{proof}

\subsection{A joint temperature--precision regime}
The preceding bounds match in a finite-temperature regime where the
inverse temperature grows with the required precision.
\begin{theorem}\label{thm:cold-minimax}
There are positive constants $c,C$ depending only on $\Cstar$ such
that, for $0<\eps\le1/16$ and every rational
$\beta\ge512/\eps$,
\begin{equation}\label{eq:cold-minimax}
 c\eps^{-3}\le\mathsf V_\beta^*(\eps)\le C\eps^{-3}.
\end{equation}
The constants are uniform in $\beta$. The upper learner has
$L_{\max}=O(\eps^{-1})$ and $O(\log^2(1/\eps))$ copies.
\end{theorem}
\begin{proof}
Put $R=\lfloor\kappa/\eps\rfloor$. Since $d=1/2$,
\cref{thm:cold-upper} applies at $\beta\ge512/\eps$.
For the lower bound, $R\ge1$ and
$\log(12(R+1))\le3+R\le4R$ imply
\begin{equation}
 \kappa\beta\ge512R\ge48\log(12(R+1)).
\end{equation}
By \cref{lem:sector-probability},
$\xi_{\beta,R}\le(R+1)^{-3}$.
At $\delta=1/10$,
\cref{lem:approximate-cost} implies
$\mathbb E[V]\ge(2/5)(R+1)^3$.
Since $R+1>\kappa/\eps$, this is the lower bound.
\end{proof}

\begin{corollary}[Copies below the size bound]
\label{cor:cold-copies}
Let $0<\eps\le1/16$, $0<\delta\le1/4$ and a rational
$\beta\ge512/\eps$. Every $(\eps,\delta)$ learner on $\Cstar$ restricted to
$L\le\lfloor\kappa/\eps\rfloor$ satisfies, on each of two instances,
\begin{equation}\label{eq:cold-copy-lower}
 \mathbb E[N]\ge(1-2\delta)e^{13\kappa\beta/8}.
\end{equation}
\end{corollary}
\begin{proof}
Put $\lambda=\kappa\beta$, $R=\lfloor\kappa/\eps\rfloor$ and
$A=12(R+1)$. The temperature condition implies
$\lambda\ge48\log A$, as above. The proof of
\cref{lem:sector-probability} gives
$\xi_{\beta,R}\le(e^2/6!)A^{18}e^{-2\lambda}$.
Since $A^{18}\le e^{3\lambda/8}$ and $e^2<6!$,
$\xi_{\beta,R}\le e^{-13\lambda/8}$.
In the hybrid argument of \cref{lem:approximate-cost}, all queries have
$L\le R$. The output total variation distance is therefore at most
$\xi_{\beta,R}\mathbb E[N]$ under either instance.
Correctness makes it at least $1-2\delta$, proving the claim.
\end{proof}

Theorem~\ref{thm:cold-minimax} allows the inverse temperature to grow
as the error decreases. In this regime larger systems give polynomial
spin cost. At fixed positive error small enough for
\cref{prop:low-temperature}, restricting queries to $L\le2$ instead
forces an exponential number of copies in $\beta$.
\Cref{sec:conclusion} identifies the intermediate temperature regime
whose spin complexity remains to be determined.

\section{Estimation at Fixed Temperature}
\label{sec:thermal}
At fixed positive temperature, copies of one-site and eight-site Gibbs
states determine the bulk interaction up to a scalar and a boundary
term. Matching local expectations is used for Hamiltonian identification in
Anshu, Arunachalam, Kuwahara and Soleimanifar~\cite[full paper, Proposition~4]{AnshuManyBody2021}; recovery from eigenstates is studied
by Qi and Ranard~\cite[Section~2.2]{QiRanard2019}.
We use logarithmic inversion and linear recovery, then locate the
transition by classical finite-energy comparisons.

\subsection{Recovering the bulk interaction}
Fix a normalized partial trace: for an operator $X$ on a finite tensor
product, $\mathcal E_S(X)$ traces out the complement and divides by its
dimension. This is a unital completely positive contraction in operator
norm on Hermitian operators. Separate the scalar, one-site and two-site
parts of each local term:
\begin{equation}
 \begin{aligned}
 h_0&=c_0I+g_0,\\
 h_i&=c_iI+A_i\otimes I+I\otimes B_i+G_i,
       \qquad i=1,2,3.
 \end{aligned}
\end{equation}
Here $g_0,A_i,B_i$ are traceless and both normalized one-site partial
traces of $G_i$ vanish. All these operators are defined by the stated
partial traces. Put
\begin{equation}
 g_\infty=g_0+\sum_{i=1}^3(A_i+B_i).
\end{equation}
Define the Hamiltonian with these on-site and two-site terms by
\begin{equation}\label{eq:reconstructed-form}
 \overline H_L(\phi)=\sum_x(g_\infty-\phi P)_x
       +\sum_{i=1}^3\sum_{x,x+e_i\in\Lambda_L}(G_i)_{x,x+e_i}.
\end{equation}
At each interior site, the original incident edges contribute every
$A_i$ and $B_i$ once. Only boundary sites miss such a contribution.
The two finite Hamiltonians therefore differ by a boundary operator of
norm $O_{J,J_b}(L^2)$ and a scalar. Their limiting energies differ by the
field-independent constant $c_0+c_1+c_2+c_3$. Hence their energy
secants and transition location agree.

\begin{lemma}[Reconstruction of the local terms]
\label{lem:bulk-recovery}
The trace-free parts of $K_1$ and $K_2$ determine $g_\infty,G_1,G_2,G_3$
by a bounded linear map whose norm depends only on $q$. If both inputs
have operator-norm error at most $\nu$, each recovered edge term and
the recovered on-site term have error at most $3\nu$. Their total
error in energy density is at most $12\nu$.
\end{lemma}
\begin{proof}
The trace-free part of $K_1$ is $g_0$. On the cube of side two, let
$C_x=\mathcal E_{\{x\}}(K_2-q^{-8}\operatorname{tr}(K_2)I)$.
Every vertex has one neighbour in each coordinate direction. Thus
$C_x=g_0+\sum_i D_{i,x}$, with $D_{i,x}$ equal to $A_i$ or $B_i$
according to the endpoint of its directed edge. Averaging over all eight
vertices, write $\overline C=\frac18\sum_{x\in\Lambda_2}C_x$.
Half the vertices contribute $A_i$ and half contribute $B_i$ in each
direction. Therefore the bulk one-site term is
\begin{equation}
 g_\infty=2\overline C-g_0.
\end{equation}
For one edge $\{x,y\}$ in direction $i$, the two-site term with vanishing
partial traces is
\begin{equation}
 G_i=\mathcal E_{\{x,y\}}(K_2-q^{-8}\operatorname{tr}(K_2)I)
       -C_x\otimes I-I\otimes C_y.
\end{equation}
Every other two-site connected term has zero partial trace in this
expression. Contractivity gives error at most $\nu$ in each $C_x$ and in
$\overline C$, at most $3\nu$ in $g_\infty$, and at most $3\nu$ in each
$G_i$. There are at most one edge per site in each of three directions.
The operator-norm perturbation of the ground energy per site is hence
at most $3\nu+3(3\nu)=12\nu$.
\end{proof}

\subsection{Stable inversion of a constant-size Gibbs state}
At field $\phi=0$,
$\|H_{u,1}(0)\|\le J+J_b$ and
$\|H_{u,2}(0)\|\le20J+4J_b$.
A common spectral lower bound for the two density matrices is
\begin{equation}
 \mu_0=q^{-8}\exp[-2\beta(20J+4J_b)].
\end{equation}
Choose a positive rational $\mu\in[\mu_0/4,\mu_0/2]$ for all
subsequent arithmetic.
If a Hermitian estimate $\widehat\rho$ has
$\|\widehat\rho-\rho\|\le t\le\mu/2$, it is positive definite.
Apply functional calculus to the integral representation of the
logarithm in Fawzi, Saunderson and
Parrilo~\cite[full paper, Section~2, Eq.~(6)]{FawziLog2019}.
The substitution $s=(1-\tau)/\tau$ for the integral parameter $\tau$ gives
\begin{equation}\label{eq:log-resolvent}
 \begin{aligned}
 \log\widehat\rho-\log\rho
 &=\int_0^\infty\bigl[(\rho+sI)^{-1}\\
 &\hspace{45pt}-(\widehat\rho+sI)^{-1}\bigr]ds.
 \end{aligned}
\end{equation}
The resolvent identity in \cref{eq:log-resolvent} bounds the integrand
norm by $t(s+\mu/2)^{-2}$. Integrating proves
\begin{equation}
 \|\log\widehat\rho-\log\rho\|\le\frac{2t}{\mu}.
\end{equation}

Writing $\Pi_0(X)=X-\operatorname{tr}(X)I/\dim(X)$, the identity
\begin{equation}
 \Pi_0(K_L)
 =-\beta^{-1}\Pi_0(\log\rho^\beta_{u,L}(0))-\Pi_0(B_L)
\end{equation}
turns state error $t\le\mu/2$ into operator-norm error at most
$4t/(\beta\mu)$ in the traceless Hamiltonian.

For an explicit rational estimator in dimension $D$, measure the
following $D^2$ observables on independent copies:
\begin{equation}\label{eq:coordinate-observables}
 \begin{aligned}
 A_j&=|j\rangle\langle j|, &&1\le j\le D,\\
 X_{jk}&=|j\rangle\langle k|+|k\rangle\langle j|, &&j<k,\\
 Y_{jk}&=i\bigl(|j\rangle\langle k|-|k\rangle\langle j|\bigr), &&j<k.
 \end{aligned}
\end{equation}
Each outcome belongs to $[-1,1]$. Define the sample threshold
\begin{equation}\label{eq:coordinate-sample-count}
 s_D=\frac{2D^2}{t^2}\log\frac{4D^2}{\delta}.
\end{equation}
Use an integer number of copies per observable between $s_D$ and $s_D+2$.
The empirical means define a rational Hermitian matrix $R_D$ by
\begin{equation}\label{eq:coordinate-estimator}
 (R_D)_{jk}=\begin{cases}
 \overline A_j,&j=k,\\
 (\overline X_{jk}+i\overline Y_{jk})/2,&j<k,\\
 \overline{(R_D)_{kj}},&j>k.
 \end{cases}
\end{equation}
Hoeffding's inequality~\cite[Theorem~2, Eq.~(2.6)]{Hoeffding1963}
and a union bound make every empirical mean accurate to $t/D$ with
probability at least $1-\delta/2$. Each matrix entry then has error
at most $t/D$, so $\|R_D-\rho\|\le\|R_D-\rho\|_F\le t$.
This gives $O_D(t^{-2}\log(1/\delta))$ copies. General state-copy
tomography bounds are developed in~\cite[Theorems~1--2 and
Eq.~(6)]{HaahTomography2016} and~\cite[Theorem~1]{GutaTomography2020}.

We also specify the spectral truncation at finite precision.
Let $f_\mu(x)=\max\{x,\mu/2\}$. The rational Bernstein-polynomial
construction in \cref{app:spectral-truncation} produces $\sigma_D$
with spectrum in $[\mu/2,D]$ and
$\|\sigma_D-f_\mu(R_D)\|\le t$.
On the tomography event, $t\le\mu/4$ implies
$f_\mu(R_D)=R_D$, hence $\|\sigma_D-\rho\|\le2t$.
The two sizes succeed jointly with probability at least $1-\delta$.
Logarithmic inversion and \cref{lem:bulk-recovery} then give energy-density
error at most $96t/(\beta\mu)$.
Compute the logarithms, subtract the known boundary terms and round the
recovered local terms so that these operations add at most
$d\eps/512$ to their summed operator-norm errors. Choose a rational
Hermitian $\widetilde P$ with $\|\widetilde P-P\|\le d\eps/512$.
Since $\phi\in[0,1]$, this adds at most $d\eps/512$ to the uniform
energy-density error.

\subsection{From approximate energies to a transition estimate}
The algorithm compares estimated finite energies with the true
limiting energy, using the true model's phase promise.
Let $\widehat H_L$ be \cref{eq:reconstructed-form} with the estimated
local terms and $\widetilde P$, and put
$\widehat e_L=L^{-3}\lambda_{\min}(\widehat H_L)$.
Write $\overline e_u$ for the limiting energy density of
\cref{eq:reconstructed-form}; it differs from $e_u$ by a constant.
The true interaction in \cref{eq:reconstructed-form} has $\|G_i\|\le4J$, so it has surface-energy
constant $C_*=\max\{1,12J\}$. Therefore
\begin{equation}
 |\widehat e_L(\phi)-\overline e_u(\phi)|\le\Delta+C_*/L.
\end{equation}
\begin{lemma}[Comparisons from approximate energies]
\label{lem:robust-energy-comparison}
Let $u\in\Cfield$ and $0<\eps\le1/16$. Suppose rational estimates of
the local terms obey the preceding energy bound, with
$\Delta\le d\eps/128$. The following deterministic finite-energy
search returns an estimate within $\eps$ of $\theta_u$. The largest
cube used in this classical computation has side length
$O(C_*/(d\eps))$.
\end{lemma}
\begin{proof}
The remaining steps are classical. At a search interval with midpoint
$\phi$ and width $w$, put $\gamma=w/8$ and $h=\gamma/2$. Construct
the estimated Hamiltonian on side length
$L=\lceil16C_*/(d\gamma)\rceil$. Let $\widetilde e_L$ approximate
its ground energy per site to error $\zeta=d\gamma/32$. The comparison
statistic is
\begin{equation}\label{eq:energy-search-statistic}
 \widehat s=-\frac{\widetilde e_L(\phi+h)-\widetilde e_L(\phi-h)}{2h}.
\end{equation}
Set $Z_k=1$ when $\widehat s\ge(a+b)/2$, and $Z_k=0$ otherwise,
in the search of \cref{subsec:interval-search}.
Every executed round has $\gamma>\varepsilon/4$.
The negative computed secant differs from the true negative limiting
secant by at most
\begin{equation}
 \frac{\Delta+C_*/L+\zeta}{h}
 \le\frac{d}{16}+\frac{d}{8}+\frac{d}{16}
 =\frac d4.
\end{equation}
At distance at least $\gamma$ from the transition, the true secant
lies on the prescribed side of $[a,b]$. The error bound therefore
gives the correct comparison with $(a+b)/2$, and
\cref{lem:interval-invariant} proves correctness.
\end{proof}

\begin{theorem}[Gibbs-state sample complexity]
\label{thm:thermal-upper}
Fix $q=3$, $J,J_b,d$ and a public rational $0<\beta<\infty$.
For $0<\eps\le1/16$ and $0<\delta\le1/4$, there is an
$(\eps,\delta)$ learner for $\Cfield$ using copies of Gibbs states
only at $L=1,2$ and $\phi=0$. Its copy count satisfies
\begin{equation}
 N\le C_{\beta,q,J,J_b,d}\,
              \varepsilon^{-2}\log\frac1\delta.
\end{equation}
The total number of spins is at most $8N$.
\end{theorem}
\begin{proof}
Choose
\begin{equation}\label{eq:tomography-accuracy}
 t=\min\left\{\frac\mu4,\frac{\beta\mu d\eps}{24576}\right\}.
\end{equation}
The tomography contribution to the bulk error is at most
$96t/(\beta\mu)\le d\eps/256$. Numerical local-term errors and the
approximation of $P$ contribute at most another $d\eps/256$.
Thus $\Delta\le d\eps/128$. For the fixed public parameters, $t$ is proportional to $\eps$.
Tomography therefore uses $O(t^{-2}\log(1/\delta))$ copies.
Each copy has at most eight spins. On the
joint tomography event, \cref{lem:robust-energy-comparison} returns the
required estimate. The failure probability is at most $\delta$.
\end{proof}

The constant in this bound contains inverse powers of $\mu$.
\Cref{sec:computation} gives the classical time for evaluating the
reconstructed finite energies. The bound keeps $\beta$ fixed as
$\eps$ decreases.

\section{Information Bounds for Gibbs States}
\label{sec:thermal-information}
We bound the information per spin in a product Gibbs family.
For positive-definite density matrices, relative entropy is
\begin{equation*}
 D_{\rm rel}(\rho\Vert\sigma)
 =\operatorname{tr}\rho(\log\rho-\log\sigma).
\end{equation*}
For classical laws, use the corresponding definition
\begin{equation*}
 D_{\rm rel}(\mathsf P\Vert\mathsf Q)
 =\begin{cases}
 \displaystyle\int\log\!\left(\frac{d\mathsf P}{d\mathsf Q}\right)d\mathsf P,
        &\mathsf P\ll\mathsf Q,\\
 +\infty,&\text{otherwise}.
 \end{cases}
\end{equation*}
We write $\operatorname{kl}(p,q)$ for the relative entropy between
Bernoulli laws with parameters $p$ and $q$.

\subsection{Information in the thermal copies}
Use zero bulk edge interaction, the same
$P$ and boundary as in $\Cstar$, and on-site term
\begin{equation}\label{eq:thermal-hard-instance}
 h_\theta(\phi)=\operatorname{diag}(0,\theta,\theta+c)-\phi P,
 \qquad c=1/8.
\end{equation}
For $\theta\in[1/4,3/4]$ the term before the field is added has norm at most
$7/8$, and its zero-temperature order parameter jumps from zero to one
at $\theta$. This satisfies \cref{eq:phase-promise}, so it is
a subfamily of $\Cstar$.

The Gibbs states are diagonal products. At site $x$, the event that
the spin is $B$ or $C$ has logistic parameter
\begin{equation}\label{eq:thermal-logistic}
 z_x=\beta(\phi-\theta)
   +\log\!\left(1+e^{-\beta(c+\kappa n_{\rm ext}(x))}\right).
\end{equation}
Conditional on that event, the $B/C$ distribution is independent of
$\theta$. The log-partition function $F(z)=\log(1+e^z)$ has
$0\le F''(z)\le1/4$. The use of log-partition derivatives to control
Gibbs-state identification is developed in~\cite[full paper, Sections~3.1--3.2]{AnshuManyBody2021}. Here the one-parameter calculation
is explicit: with $p_z=F'(z)$, Taylor's theorem gives
\begin{equation}\label{eq:bernoulli-divergence}
 \begin{aligned}
 \operatorname{kl}(p_z,p_{z'})
 &=F(z')-F(z)-F'(z)(z'-z)\\
 &\le\frac{(z-z')^2}{8}.
 \end{aligned}
\end{equation}
The conditional $B/C$ law cancels from relative entropy. Applying
\cref{eq:bernoulli-divergence} at each site and adding the independent
contributions yields
\begin{equation}\label{eq:thermal-relative-entropy}
 D_{\rm rel}\!\left(\rho^\beta_{\theta_0,L}(\phi)
        \middle\|\rho^\beta_{\theta_1,L}(\phi)\right)
 \le\frac{\beta^2(\theta_0-\theta_1)^2}{8}\,L^3.
\end{equation}
Because the states are diagonal in a known basis, a basis configuration
is sufficient to prepare the quantum input. Every subsequent quantum
instrument, including joint processing of different copies, is an
instance-independent channel on these configurations.

\begin{theorem}[Thermal information lower bound]
\label{thm:thermal-lower}
For $0<\eps\le1/16$ and $0<\delta\le1/4$, two instances from
\cref{eq:thermal-hard-instance} force every correct almost-surely
terminating algorithm to satisfy
\begin{equation}\label{eq:thermal-cost-lower}
 \mathbb E_{\theta_0}[V]\ge\frac{1}{8\beta^2\eps^2}\log\frac1\delta.
\end{equation}
\end{theorem}
\begin{proof}
Take $\theta_0=1/2-2\eps$ and $\theta_1=1/2+2\eps$.
A query has relative entropy at most $2\beta^2\eps^2 L^3$.
We use the stopped likelihood argument of Kaufmann, Capp\'e and
Garivier~\cite[Lemma~1; Appendix~A.1, Eq.~(18)]{KaufmannCappeGarivier2016}.
Expose the full basis configuration at each query. Conditional on
these configurations, the quantum processing is an
instance-independent channel. The query policy contributes the same
conditional probabilities under both instances; the log-likelihood
increment of a fresh configuration has conditional mean equal to its
relative entropy. For an integer horizon $T$, let $\mathsf P_i^{(T)}$
be the actual classical transcript law under $\theta_i$, where
$i\in\{0,1\}$, through the first $T$ queries. A terminated run is
padded by a fixed symbol. The conditional likelihood calculation
bounds the relative entropy of the latent configurations.
The instance-independent measurement channel contracts relative
entropy, by data processing~\cite{Lindblad1975}. Hence
\begin{equation}\label{eq:stopped-information}
 D_{\rm rel}(\mathsf P_0^{(T)}\Vert\mathsf P_1^{(T)})
 \le2\beta^2\eps^2\,
      \mathbb E_{\theta_0}\!\left[\sum_{t=1}^{\min\{N,T\}}L_t^3\right].
\end{equation}
This argument applies to the countable field and size query set.
Use the common event
$A_T=\{N\le T,|\widehat\theta-\theta_0|\le\eps\}$.
Applying data processing to this event bounds transcript relative entropy below by
$\operatorname{kl}(\Pr_{\theta_0}(A_T),\Pr_{\theta_1}(A_T))$.
For these diagonal inputs, the binary-event inequality follows from
the log-sum inequality on $A_T$ and its complement.
As $T$ increases, the first probability tends to at least $1-\delta$
and the second to at most $\delta$, because the two success intervals
are disjoint. Lower semicontinuity of binary relative entropy and
monotone convergence of the nonnegative expected cost sum give
\begin{equation*}
 2\beta^2\eps^2\mathbb E_{\theta_0}[V]\ge\operatorname{kl}(1-\delta,\delta).
\end{equation*}
For $\delta\le1/4$, the binary divergence satisfies
\begin{equation*}
 \begin{aligned}
 \operatorname{kl}(1-\delta,\delta)
 &=(1-2\delta)\log\frac{1-\delta}{\delta}\\
 &\ge\frac12\log\frac1{\sqrt\delta}\\
 &=\frac14\log\frac1\delta.
 \end{aligned}
\end{equation*}
The inequality uses $1-2\delta\ge1/2$ and
$1-\delta\ge\sqrt\delta$ on the stated range. Substitution gives
\cref{eq:thermal-cost-lower}.
\end{proof}

\subsection{The temperature dependence at bounded system size}
\label{subsec:low-temperature}
The ferromagnetic family also quantifies the sample cost of lowering
the temperature at bounded system size.
The information loss from a rarely occupied excited level appears in
Haah, Kothari and Tang~\cite[Supplementary Information,
Lemma~8.2]{HaahKothariTang2024}. Here we compute it for the
ferromagnetic family, uniformly over the learner's field choices.

\begin{proposition}\label{prop:low-temperature}
Let $0<\eps\le\kappa/4$, $0<\delta\le1/4$ and a rational $\beta>0$.
Every $(\eps,\delta)$ learner for $\Cstar$ restricted to $L\le2$
satisfies, on each of two instances,
\begin{equation}\label{eq:activated-sample-lower}
 \mathbb E[N]\ge
 \frac{e^{2\kappa\beta}}{2048\,3^8\beta^2\eps^2}
       \log\frac1\delta.
\end{equation}
Collective measurements and almost-sure stopping are allowed.
\end{proposition}
\begin{proof}
Use the ferromagnetic family with $r_0=0$ and $r_1=4\eps$.
Along the interval between these parameters, $r\le\kappa$.
For a configuration $s$, let $N_A,N_B,N_C$ count its three colours,
let $n=L^3$, and put $h=1/2-\phi$ and $g=\min\{0,hn\}$.
The exact energy decomposition gives
\begin{equation}\label{eq:c-excitation-gap}
 \begin{aligned}
 E_r(s)-g
 &=h(N_B+N_C)-g-rN_C\\
 &\quad+\kappa|\partial S_C|+\kappa e_{AB}\\
 &\ge(6\kappa/L-r)N_C\\
 &\ge2\kappa N_C.
 \end{aligned}
\end{equation}
For $h\ge0$ the first two terms on the right are nonnegative;
for $h<0$ their difference is $-hN_A\ge0$.
The uniform $A$ or $B$ state has energy $\min\{0,hn\}$, so the
partition function is at least its Boltzmann weight. There are at most
$3^8$ configurations. Consequently
\begin{equation}
 \operatorname{Var}_r(N_C)
 \le\mathbb E_r[N_C^2]
 \le64\,\Pr_r\{N_C>0\}
 \le64\,3^8 e^{-2\kappa\beta}.
\end{equation}
The second derivative of $\log Z_r$ is
$\beta^2\operatorname{Var}_r(N_C)$. Taylor's theorem bounds the
relative entropy between the two Gibbs states of any allowed query by
$512\,3^8\beta^2\eps^2e^{-2\kappa\beta}$.
Applying the stopped likelihood argument of \cref{thm:thermal-lower}
with unit cost per copy proves \cref{eq:activated-sample-lower}.
Interchanging the two instances gives the bound under each one.
\end{proof}

The choice $\mu\in[\mu_0/4,\mu_0/2]$ in the Gibbs-state algorithm also
makes its temperature dependence explicit. Constants $C_1,C_2>0$
depending only on $\Cstar$ give
\begin{equation}\label{eq:temperature-dependent-upper}
 N\le C_1 e^{C_2\beta}\bigl(1+(\beta\eps)^{-2}\bigr)
                    \log\frac1\delta.
\end{equation}
The sample count is $O(t^{-2}\log(1/\delta))$.
By \cref{eq:tomography-accuracy},
\begin{equation}
 t^{-2}\le\mu^{-2}\left(16+\frac{24576^2}{\beta^2d^2\eps^2}\right).
\end{equation}
Together with \cref{prop:low-temperature}, this shows an exponential
worst-case dependence on inverse temperature at bounded system size.

\section{Classical Computation after the Measurements}
\label{sec:computation}
We account for classical processing after receiving the state copies.
Hamiltonian-learning results such
as~\cite[full paper, Theorem~1.1]{BakshiHamiltonian2026}
and~\cite[Theorem~6]{Narayanan2025} state both quantities for coefficient
recovery. The fixed-temperature learner estimates two constant-dimensional
density matrices, recovers rational local terms in
\cref{eq:reconstructed-form}, and computes finite-volume ground energies.
At a search
margin $\gamma$, the required side length is
$L=\lceil16C_*/(d\gamma)\rceil$ and the energy-density accuracy is
$d\gamma/32$. The matrix dimension is $q^{L^3}$.

For rational local terms, exact positive-definiteness decisions and
bisection approximate the least eigenvalue in polynomial bit complexity
in that dense matrix dimension and the entry bit length. The proof in
\cref{app:description-proof} gives the elimination and bit-length bounds.
Let $B_{\rm acc}$ be the bit length of the rational accuracy and
confidence inputs. They can first be rounded down to dyadic values
within a factor two. For fixed physical parameters, the classical time is
\begin{equation}\label{eq:description-time}
 T(\eps,\delta)\le\operatorname{poly}(B_{\rm acc})+q^{O(\eps^{-3})}
        \operatorname{poly}\!\left(\log\frac1\eps,\log\frac1\delta\right).
\end{equation}
The ground-state and larger-system Gibbs learners
(\cref{target:active-learning,thm:cold-upper}) use random-site
selection, binary measurements, count updates and rational interval
arithmetic. Their expected classical processing is polynomial in
$B_{\rm acc}$. Thus the fixed-temperature sample bound is attained
by a learner with exponential classical time, while the local-measurement
learners have polynomial classical time.

\subsection{Classical descriptions}
The same energy search also gives an effective approximation from a
finite description of the interaction. Here the data are the local
matrices themselves.
\begin{corollary}\label{cor:description-algorithm}
Suppose $h_0,h_1,h_2,h_3$ and $P$ are rational Hermitian matrices,
and the interaction satisfies \cref{eq:phase-promise}. For fixed
$q,J,d$, a deterministic algorithm given these matrices and
$0<\eps\le1/16$ returns an estimate within $\eps$ of $\theta_u$ in time
\begin{equation}
 q^{O(\eps^{-3})}\operatorname{poly}(B_{\rm loc},B_{\rm acc}),
\end{equation}
where $B_{\rm loc}$ is the bit length of the local matrices.
\end{corollary}
\begin{proof}
Omit the boundary term, which leaves the limiting energy unchanged.
The block estimate gives energy-density error at most $3J/L$.
Use the finite-energy comparisons with interaction error zero and
$C_*=\max\{1,3J\}$. The same interval invariant gives the estimate.
The rational eigenvalue procedure in \cref{app:description-proof}
provides the time bound.
\end{proof}
The derivative-gap promise thus supplies an effective approximation
rate for this description problem. At fixed positive precision and
fixed $q,J,d$, the time is polynomial in the input bit length.
Watson and
Bausch~\cite[Definition~10 and Theorem~19]{WatsonBauschCritical2021}
study critical-point complexity under a local--global phase promise.
Their construction has phase-encoded two-site
terms~\cite[Theorem~36, item~2]{WatsonBauschCritical2021}.
Here the field couples linearly to a fixed on-site observable, whose
expectation is a supergradient of the energy. This relation gives the
explicit approximation rate in the corollary.

\section{Discussion}
\label{sec:conclusion}
The size constraint in \cref{cor:finite-size-separation} has an
exponential effect on copy complexity even when the learner controls
the field and uses collective measurements. The common surface field
preserves identical small-system ground states and suppresses the
thermal configurations carrying the transition information.
Boundary counting controls this suppression uniformly over the allowed
queries; the hybrid argument then handles quantum memory and stopping.

The upper learner uses the conjugate relation between the known field
and the order parameter. Free-energy concavity gives local comparisons,
and geometric growth of the queried cubes gives the optimal
$\Theta(\eps^{-3})$ total spin count for $\beta\ge512/\eps$.
At fixed positive temperature, logarithmic inversion of constant-size
states gives the different statistical rate $\Theta_\beta(\eps^{-2})$,
with exponential classical postprocessing.

Determining $\mathsf V_\beta^*(\eps)$ for
$1\ll\beta\ll\eps^{-1}$ requires a joint analysis of thermal
information and finite-size energy error.
\Cref{prop:low-temperature,lem:approximate-cost} give bounded-size
and all-size lower bounds for this question. A second problem is to
determine the size--sample tradeoff with periodic boundaries or zero
surface field, where the available finite-system states are different.

\bibliographystyle{ACM-Reference-Format}
\bibliography{references}
\appendix
\section{Ground-State Learning by Local Measurements}
\label[appendix]{sec:main-results}

\subsection{A finite-size bound for the order parameter}
\label{subsec:resolution}
Let $J$ bound each nearest-neighbour interaction. The boundary norm
is at most $J_bL^2$. Put $C=\max\{1,3J+J_b\}$.
The energy density $e_{u,L}(\phi)=L^{-3}\lambda_{\min}(H_{u,L}(\phi))$
has a thermodynamic limit $e_u(\phi)$ satisfying
\begin{equation}\label{eq:surface-bound}
 \sup_{u,\phi}|e_{u,L}(\phi)-e_u(\phi)|\le C/L.
\end{equation}
For block tiling, see Aharonov and
Irani~\cite[full paper, Lemma~3.2]{AharonovIraniThermodynamic2025}.
The proof in \cref{app:proofs} counts the interaction terms between
cubic blocks. It applies to noncommuting bulk interactions.

\begin{theorem}[Finite-size order-parameter bound]
\label{target:resolution}
For every instance in $\Cfield$, every $L\ge1$, and every $\gamma>0$,
\begin{equation}\label{eq:resolution}
 m_{u,L}(\phi)\in
 \begin{cases}
 [0,a+4C/(L\gamma)],&\phi\le\theta_u-\gamma,\\
 [b-4C/(L\gamma),1],&\phi\ge\theta_u+\gamma.
 \end{cases}
\end{equation}
\end{theorem}
\begin{proof}
The variational principle makes $-m_{u,L}(\phi)$ a supergradient of the
concave function $e_{u,L}$, including at a ground-state degeneracy.
Uniform convergence in \cref{eq:surface-bound} carries this property to
the limiting order parameter. Thus $e_u'$ belongs to $[-a,0]$ almost
everywhere to the left of $\theta_u$, and to $[-1,-b]$ to the right.
For $\phi\le\theta_u-\gamma$, compare the supergradient with the forward
secant of length $h=\gamma/2$. The limiting secant is at least $-a$,
and the energy error changes it by at most $2C/(Lh)$. Hence
$-m_{u,L}(\phi)\ge-a-4C/(L\gamma)$. The backward secant proves the
other case. The position $\theta_u\in[1/4,3/4]$ keeps these comparison
intervals inside the control domain. Details appear in \cref{app:proofs}.
\end{proof}

\subsection{The learning algorithm}
\label{subsec:learner}
Use the interval search of \cref{subsec:interval-search}. At round $k$,
query
$L_k=\lceil16C/(d\gamma_k)\rceil$.
Outside the $\gamma_k$-neighbourhood of the transition, the mean of
the local measurement is separated from $(a+b)/2$ by at least
$g=d/4$.

Allocate failure probability $\delta_k=\delta 2^{k-K}$ to round $k$.
Take an integer number $n_k$ of independent copies between
$2g^{-2}\log(2/\delta_k)$ and this quantity plus two, using a rational
upper approximation to the logarithm.
If $\overline Y_k$ denotes their mean, set $Z_k=1$ when
$\overline Y_k\ge(a+b)/2$ and $Z_k=0$ otherwise.

\begin{theorem}[Ground-state upper bound]
\label{target:active-learning}
For $0<\eps\le1/16$ and $0<\delta\le1/4$, the algorithm belongs to
$\mathfrak A^{\infty}_{\eps,\delta}(\Cfield)$. A universal constant $A$ gives
\begin{equation}\label{eq:volume-upper}
 \sup_{u\in\Cfield}\E_u[V]
 \le\frac{AC^3}{d^5\eps^3}\log\frac1\delta.
\end{equation}
Its maximum side length is $O(C/(d\eps))$. Each field value has a
dyadic encoding of $O(\log(1/\eps))$ bits. The number of copies is
\begin{equation}
 N=O\!\left(d^{-2}\left[K\log\frac1\delta+K^2\right]\right).
\end{equation}
\end{theorem}
\begin{proof}
Conditional on each history, Hoeffding's
inequality~\cite[Theorem~2, Eq.~(2.6)]{Hoeffding1963} bounds
$\Pr\{|\overline Y_k-m_{u,L_k}(\phi_k)|\ge g/2\}$ by $\delta_k$.
Their sum is less than $\delta$. On the complementary event, every
comparison farther than $\gamma_k$ from the transition has the correct
sign, and \cref{lem:interval-invariant} preserves containment. The final
width is at most $2\eps$.

For cost, write $j=K-1-k$. Then
$L_k\le128C(5/8)^j/(d\eps)$ and
$n_k\le32[(\log(2/\delta)+(j+1)\log2)]/d^2+2$.
The sum of $n_kL_k^3$ is bounded by convergent geometric sums, proving
\cref{eq:volume-upper}. These sums and the exact control representation
are given in \cref{app:proofs}.
\end{proof}
The allocation assigns more observations to earlier, smaller systems.
Since system volumes grow geometrically, the aggregate bound remains
$O(\eps^{-3}\log(1/\delta))$.

\section{Additional Proofs}
\label[appendix]{app:proofs}

\subsection{The thermodynamic energy density}
We use the block-tiling method for thermodynamic energy approximation;
see Aharonov and Irani~\cite[full paper, Lemma~3.2]{AharonovIraniThermodynamic2025}.
The following calculation gives the constants for three-dimensional
cubes, on-site terms and the prescribed boundary operator.
First remove $B_L$ and write $f_L$ for the ground energy per site of the
bulk Hamiltonian. Tile an $mL$-cube by $m^3$ cubes of side $L$.
Removing the $3m^2(m-1)L^2$ interactions crossing their boundaries produces
independent blocks. Operator-norm perturbation gives
\begin{equation}\label{eq:block-bound}
 |f_{mL}(\phi)-f_L(\phi)|
 \le\frac{3J}{L}\left(1-\frac1m\right).
\end{equation}
For a general side length $n$, tile $\lfloor n/L\rfloor^3$ blocks and
leave the remaining boundary slabs. Their total volume is $O(n^2L)$;
the on-site norm is uniformly bounded by $J+1$ and the edge norm by $J$.
Removing their interactions and bounding their energies therefore yields
$|f_n-f_L|\le3J/L+O_J(L/n)$, uniformly in $u$ and $\phi$.
The limsup and liminf of $f_n$ differ by at most $6J/L$ for each $L$,
so they agree. Taking $n\to\infty$ gives $|f_L-e_u|\le3J/L$.
Finally, $\|B_L\|/L^3\le J_b/L$ proves \cref{eq:surface-bound}.

\subsection{Concavity and the order parameter}
For any ground-state density matrix at $\phi$, the variational principle
gives $e_L(\phi+h)\le e_L(\phi)-hm_L(\phi)$.
This includes the normalized ground-space projector. Passing to the
limit at points away from $\theta$ puts $-m_u(\phi)$ in the
superdifferential of $e_u$. The function $e_u$ is concave and
one-Lipschitz, hence absolutely continuous. At its differentiability
points, its derivative equals $-m_u$. Integrating the phase promise
bounds the forward secants on the left by $-a$ from below, and the
backward secants on the right by $-b$ from above. The two energy errors
in each secant sum to at most $2C/L$. Taking step $\gamma/2$ proves
\cref{eq:resolution}.

\subsection{Counting spins and encoding the field values}
\label[appendix]{app:spin-count}
Let $\rho=125/512$ and put
$\Psi_\delta=\log(2/\delta)/(1-\rho)+\log2/(1-\rho)^2$.
The bounds in the main proof give the explicit estimate
\begin{equation}\label{eq:explicit-volume}
 V\le\frac{2^{21}C^3}{d^3\eps^3}
 \left[\frac{32\Psi_\delta}{d^2}
             +\frac{2}{1-\rho}\right].
\end{equation}
Indeed, $\sum_{j\ge0}\rho^j=(1-\rho)^{-1}$ and
$\sum_{j\ge0}(j+1)\rho^j=(1-\rho)^{-2}$.
Since $\delta\le1/4$, this is \cref{eq:volume-upper} for a universal
constant. Summing $n_k$ instead gives the stated copy count.
Use denominator $Q=2^{3K+2}$ for interval endpoints. At round $k$, the
width numerator is $5^k2^{3K+1-3k}$, so both midpoint and one-eighth
width are exact integers in these coordinates. Thus every field value is
dyadic with $O(K)$ bits in each learner.

\subsection{Rational spectral truncation}
\label[appendix]{app:spectral-truncation}
The entries of $R_D$ in \cref{eq:coordinate-estimator} are rational,
and $\|R_D\|\le\|R_D\|_F\le D$ on every outcome. Put
$T_D=(R_D+DI)/(2D)$, whose spectrum lies in $[0,1]$, and set
$m=\lceil D^2/t^2\rceil$. Define the rational matrix
\begin{equation}\label{eq:bernstein-floor}
 \sigma_D=\sum_{k=0}^{m}
   f_\mu(2Dk/m-D)\binom{m}{k}T_D^k(I-T_D)^{m-k}.
\end{equation}
Let $K\sim\operatorname{Bin}(m,x)$ for $x\in[0,1]$.
Its value is $\mathbb E f_\mu(2DK/m-D)$.
The function $f_\mu$ is one-Lipschitz. Cauchy--Schwarz therefore gives
\begin{equation}
 \left|\mathbb E f_\mu(2DK/m-D)-f_\mu(2Dx-D)\right|
 \le2D\sqrt{\frac{x(1-x)}{m}}\le\frac{D}{\sqrt m}\le t.
\end{equation}
The spectral theorem gives the same operator-norm error for $R_D$.
Each coefficient value belongs to $[\mu/2,D]$, so every eigenvalue of
$\sigma_D$ belongs to that interval as well. This constructs the
required rational positive matrix on every outcome. Matrix powers,
binomial coefficients and rational sums use polynomial bit complexity
in $m$, $D$ and the input bit length. At fixed physical parameters,
this cost is polynomial in $1/\eps$.

\subsection{Bit complexity of the finite-energy computation}
\label[appendix]{app:description-proof}
We implement the finite-energy approximation used in
\cref{sec:thermal,sec:computation}. Put $D_L=q^{L^3}$ and $A=\widehat H_L(\phi)/L^3$. For a rational matrix $X$, let
$\|X\|_{\rm ent}=\sum_{i,j}(|\operatorname{Re}X_{ij}|+|\operatorname{Im}X_{ij}|)$,
which bounds its operator norm. From the reconstructed rational terms
compute the rational bound
\begin{equation}
 \widehat G=1+\|\widetilde P\|_{\rm ent}+\|\widehat g_\infty\|_{\rm ent}
                  +\sum_i\|\widehat G_i\|_{\rm ent}.
\end{equation}
The spectrum of $A$ lies in $[-\widehat G,\widehat G]$ on every
measurement outcome. The imposed spectral floor bounds these local-term
norms by a constant depending on the fixed public parameters.
For rational $t$, $A-tI$ is positive definite exactly when
$t<\lambda_{\min}(A)$. This predicate is decided by rational Schur
elimination. At a positive first pivot $a$, the block quadratic form
has the identity
\begin{equation}\label{eq:schur-quadratic}
 \begin{pmatrix}x\\z\end{pmatrix}^{\!*}
 \begin{pmatrix}a&b^*\\b&D_0\end{pmatrix}
 \begin{pmatrix}x\\z\end{pmatrix}
 =a\left|x+\frac{b^*z}{a}\right|^2
   +z^*\left(D_0-\frac{bb^*}{a}\right)z.
\end{equation}
Thus positivity reduces recursively to the Schur complement; a
nonpositive pivot certifies failure of positive definiteness. This
also handles a candidate $t$ equal to the least eigenvalue.
Bisection needs $O(\log(\widehat G/\zeta))$ such decisions.

Let $B_{\rm in}$ be the total bit length of the rational reconstructed
local terms. Their common denominator has bit length
$O(B_{\rm in})$. Forming $A-tI$ increases entry bit length
by $O(\log(1/\eps)+\log L)$, since the lattice sums repeat the same local
coefficients. Write this bit bound as $H$. After clearing denominators,
a $k$ by $k$ minor has magnitude at most $k!2^{kH}$ by the determinant
expansion. Its bit length is therefore $O(k(H+\log k))$.
Schur elimination entries are ratios of bordered minors, by
Sylvester's identity; see Bareiss~\cite[Section~I, Eqs.~(3)--(8)]{Bareiss1968}.
The displayed minor bound therefore shows that exact reduced rational
arithmetic uses a polynomial number of bits and operations
in $D_L$ and $H$. Complex rational Hermitian matrices can equivalently
be represented by their real symmetric block matrices of dimension
$2D_L$. Both representations give a polynomial dense-matrix bit cost.
There are $O(\log(1/\eps))$ search rounds. For each fixed
tomographic dimension $D_0\in\{3,3^8\}$, the logarithm of $\sigma_{D_0}$
can be approximated with polynomially many arithmetic steps in
$\log(1/\eps)$: after scaling by $D_0$, the spectrum lies in
$[\mu/(2D_0),1]$, and the
series $\log(I-X)=-\sum_{k\ge1}X^k/k$ has geometric ratio at most
$1-\mu/(2D_0)$. Scalar logarithms use range reduction to $[1,2]$
and the corresponding convergent series; rational upper bounds specify
all sample counts. The rational local terms have bit length polynomial
in $1/\eps$, the count bit lengths and the accuracy input length.
Spectral truncation and tomography take polynomial time in $1/\eps$
at fixed parameters. The dense finite-energy computation dominates
these steps and gives \cref{eq:description-time}.

\end{document}